\documentclass{aa}

\usepackage{graphics}
\usepackage{times}

\newcommand{\logt}{\mbox{$\log(t/{\rm yr})$}}

\newcommand{\logL}{\mbox{$\log(L/L_{\odot})$}}

\newcommand{\beq}{\begin{equation}}
\newcommand{\eeq}{\end{equation}}
\newcommand{\beqa}{\begin{eqnarray}}
\newcommand{\eeqa}{\end{eqnarray}}
\newcommand{\benu}{\begin{enumerate}}
\newcommand{\eenu}{\end{enumerate}}
\newcommand{\bite}{\begin{itemize}}
\newcommand{\eite}{\end{itemize}}
\newcommand{\bdes}{\begin{description}}
\newcommand{\edes}{\end{description}}
\newcommand{\ud}{{\rm d}}

\begin{document}

%	\thesaurus{06(08.05.3, 08.09.3, 08.22.3, 08.01.1) }

        \title{Zero-metallicity stars}
	\subtitle{II. Evolution of very massive objects with mass loss}

	\author{P. Marigo$^1$, C. Chiosi$^1$, R.-P. Kudritzki$^2$ }
	\institute{$^1$ Dipartimento di Astronomia, Universit\`a di Padova,
	Vicolo dell'Osservatorio 2, I-35122 Padova, Italy \\
	$^2$ Institute for Astronomy, University of Hawaii, 
	2680 Woodlawn Drive, Honolulu, HI 96822                        
		}

	\titlerunning{Evolution of zero-metallicity VMO with mass loss }
	\authorrunning{P. Marigo et al.}

	\offprints{P. Marigo \\ e-mail: marigo@pd.astro.it}

	\date{Received September 19, 2002 / Accepted November 26, 2002}

%%%%%%%%%%%%%%%%%%%%%%%%%%%%%%%%%%%%%%%%%%%%%%%%%%%%%%%%%%%%%%%%
	\abstract{
We present evolutionary models of zero-metallicity very massive 
objects, with initial masses in the range 120 $M_{\odot}$ -- 1000 $M_{\odot}$,
covering their quiescent evolution up to central carbon ignition.
In the attempt of exploring the possible occurrence of 
mass loss by stellar winds, calculations are carried out with
recently-developed formalisms for the  mass-loss rates driven 
by radiation pressure (Kudritzki 2002) and stellar rotation 
(Maeder \& Meynet 2000).
The study completes the previous analysis 
by Marigo et al. (2001) on the constant-mass evolution of primordial stars.
Our results indicate that  
radiation pressure 
(assuming a minimum metallicity $Z = 10^{-4}\times Z_{\odot}$)
is not an efficient driving force of mass loss,
except for very massive stars with $M \ga 750 \, M_{\odot}$.
On the other hand, stellar
rotation might play a crucial role in triggering  powerful stellar winds, 
once the $\Omega\Gamma$-limit is approached.
However, this critical condition of intense mass loss can be
maintained just for short, as the loss of angular momentum 
due to mass ejection quickly leads to the spinning down of the star.
As by-product to the present work, the wind chemical yields from 
massive zero-metallicity stars are presented. The
helium and metal enrichments, and the resulting $\Delta Y/\Delta Z$ ratio 
are briefly discussed.
   }

   \maketitle
%  
%
%%%%%%%%%%%%%%%%%%%%%%%%%%%%%%%%%%%%%%%%%%%%%%%%%%%%%%%%%%%%%%%%%%%%%%%%%%%

\section{Introduction}
\label{sec_intro}

The first generation of stars ever born in the Universe (Population III)
is assigned a role of paramount importance for several astrophysical 
issues. 
In particular, the formation of a primeval population of 
very massive objects (VMOs, with initial masses in the range 
$10^2-10^5 \, M_{\odot}$) was invoked in the '80s to 
explain several questions, such as the
observed metallicities of Population II stars, the primordial
helium abundance,  
the re-ionisation of cosmic matter 
after the Big-Bang, the missing mass in clusters of galaxies and
galactic haloes, the G-dwarf problem (see Carr et al. 1984 for an
extensive review). 
In those years a few evolutionary calculations of zero-metallicity 
VMOs were carried out (e.g. Bond et al. 1984, El Eid et al. 1983, 
Ober et al. 1983, Klapp 1984ab), 
but afterwards the scientific production
on the evolution of the first stars dropped off.  

The recently renewed interest in Pop-III stars,
essentially driven by the nowadays flourishing 
of observations at very low metallicity and/or high redshift
(refer to e.g. the proceedings of the ESO symposium edit by Weiss
et al. (2000); see also Kudritzki et al. (2000), Bromm et al. (2001),
Schaerer (2002), Panagia (2002) for 
extensive analyses on the expected observable properties of primordial 
stellar populations) 
has again stimulated  the calculation of stellar structures 
made evolve with initial metal-free chemical composition
(e.g. Cassisi et al. 2001; Marigo et al. 2001 and references therein).

Marigo et al. (2001; thereinafter Paper I) presented an extensive study of the
evolutionary properties of zero-metallicity stars over
a wide range of initial masses  ($0.7 \, M_{\odot} \la M \la 100 M_{\odot}$).
In that work -- to which the reader is referred for all the details -- 
we calculated all stellar tracks at constant mass.

Our study on Pop-III stars is now extended to very massive objects (VMOs),
with initial masses in the range $120-1000\, M_{\odot}$,
that is chosen not only for continuity with Paper I, but also
in consideration of current theoretical indications that the primordial
initial mass function  (IMF) might have peaked in the (very) high-mass
domain (with $M \ga 100 \, M_{\odot}$; see e.g. 
Bromm et al. 1999, 2002; Nakamura \& Umemura 2001; Abel et al. 2000).

To this aim we calculate evolutionary models for zero-metallicity VMOs 
that cover the major core-burning phases, extending from 
hydrogen to carbon ignition. With aid of available analytic formalisms, 
we address the question of the possible occurrence of mass loss 
via stellar winds, which is 
an important, but still problematic, aspect of stellar evolution
at zero metallicity.

In massive stars with ``normal'' chemical composition the principal 
driving force  
resides in the capability of metallic ions, present in the atmosphere,
to absorb radiative momentum and transfer it to the outermost layers, 
that can be then accelerated beyond their escape velocity 
(Castor et al. 1975; Pauldrach et al. 1986). 
In a gas composed only of hydrogen and helium, like the primordial one,  
the lack of metallic ions sets the first important difference 
and the question arises:
Is the radiative acceleration due to the H and He lines strong 
enough to trigger significant mass loss ?
Furthermore,  may other possible processes -- e.g.
related to pulsation instability and stellar rotation --  
be efficient mechanisms to drive mass loss from Pop-III massive stars ? 

Some of these questions have already been addressed by other investigations,
others still deserve to be quantitatively analysed and discussed.
There are few studies in the literature 
that present evolutionary models  of very massive zero-metallicity stars 
with mass loss  during the pre-supernova phases.
In the very massive domain ($500-1000\, M_{\odot}$) 
Klapp (1983, 1984) carried out evolutionary calculations by adopting a 
simple empirical law for mass-loss (Barlow \& Cohen 1977) -- based 
on Galactic observations -- that
linearly scales with the stellar luminosity.
On the basis of 
semi-analytical models of very massive objects 
(with masses $10^2 -10^5\, M_{\odot}$), Bond et al. (1984)
argued that the same kind of dynamical instability arising 
in the H-burning shell of Population I models -- which should cause the 
ejection of the entire  envelope --  might also affect
Population III VMOs, depending on the actual
abundance of the CNO catalysts in the H-shell.
The evolution of zero-metallicity stars 
(with masses $80 - 500\, M_{\odot}$) during  the nuclear phases of H- 
and He-burning was calculated by El Eid et al. (1983) with the 
adoption of a semi-empirical relation to derive  the mass loss 
rate (Chiosi 1981).
Recently Baraffe et al. (2001, see also Heger et al. 2001) 
performed a linear stability analysis
on metal-free very massive models (with masses $120-500\, M_{\odot}$),
and  investigated the related possibility that mass loss may characterise
the pulsation-unstable stages. 
Heger \& Woosley (2002) calculated the 
 evolution and nucleosynthesis of quite massive Population III stars
(with masses $\sim 100 - 300\, M_{\odot}$) including the 
final fate of the supernova event, 
but no mass-loss is assumed to occur during the hydrostatic phases
of major nuclear burnings.  

In the present work we attempt to further explore the issue
of mass loss in primordial conditions, 
by carrying out new evolutionary calculations 
of  zero-metallicity VMO  with  
recently developed mass-loss formalisms, related to 
radiation pressure and stellar rotation.

The paper is organised as follows.
Section \ref{sec_evolcalc} introduces 
new evolutionary calculations for zero-metallicity VMOs, 
having initial masses of 120, 250, 500, 750, and 1000 $M_{\odot}$.
Mass loss is included with  the adoption  of recent analytic
formalisms, namely: Kudritzki (2002) for radiation-driven mass loss, 
and Maeder \& Meynet (2000) for the additional effect of stellar rotation.
General characteristics of the models are analysed in relation to
energetics, nuclear lifetimes, internal structure, 
surface and chemical properties. The efficiency of mass loss is discussed 
on the basis of the adopted formalisms.
The corresponding predictions for the chemical yields, ejected via
stellar winds, are given in Sect.~\ref{sec_yields}.
In particular, the expected helium and metal enrichment, produced by 
a hypothetical burst of Pop-III star formation, is derived and
discussed as a function of the involved parameters.
In Sect.~\ref{sec_concl} we express a few concluding remarks. 
Finally, stellar isochrones for very young ages are presented in 
Sect.~\ref{sec_tableisoc}. 

\section{Evolutionary calculations}
\label{sec_evolcalc}
We have carried out evolutionary calculations of zero-metallicity 
very massive objects (VMO) 
with initial masses of 120, 250, 500, 750, and 1000 $M_{\odot}$.
All input physics are the same as those employed in Paper I -- to which
the reader is referred for all details -- except for the applied 
overshoot scheme. Overshooting is allowed from the borders of the convective
cores with the parameter $\Lambda_{\rm c} = 0.5$
(following Bressan et al. 1981), whereas the classical
Schwarzschild criterion is otherwise applied (i.e. convective shells
and outer envelope). 
The evolution is followed starting from the zero-age main sequence
(MS) up to the ignition of central carbon, i.e. covering 
the major stages of hydrogen and helium burning.

In order to estimate the efficiency of mass-loss via
stellar winds in primordial VMOs, 
we adopt recently-developed analytic formalisms
(Kudritzki 2002; Maeder \& Meyent 2000), 
whose basic ingredients are summarised in the next sections.
 
\subsection{Radiation-driven mass-loss rates}
\label{ssec_rad}
The role of radiation pressure 
-- on the resonance lines of metallic ions like CIII, NeII, OII -- 
in driving mass loss from massive stars has been a firm result 
of theoretical astrophysics since the pioneer works by 
Lucy \& Solomon (1970), and Castor et al. 
(1975; CAK). In short, the crucial term determining the efficiency of mass loss
 by line driven winds is the  radiative line acceleration, which is 
expressed in units of  the Thomson acceleration via the so-called force 
multiplier. This latter, in turn, is  suitably parametrised as
  $M[t(r)] = k t^{-\alpha}n^{\delta}$, 
where $t(r)$ -- the Thomson optical depth parameter -- is 
proportional to the number density of electrons,  the Thomson cross-section, 
and the ratio between the local velocity of thermal  motion to the 
spatial gradient of the stellar wind outflow velocity field; $k$ 
represents the normalisation integral over the contribution of all
photon absorbing spectral lines with different line strength; $n$ is the 
ratio of the 
local number density of electrons to the local geometrical dilution factor 
of the radiation field; and finally $\alpha$ and $\delta$ are 
suitable exponents describing the optical depth and density dependence
of the radiative line force
(see CAK, Abbott 1982a, Pauldrach et al. 1986, Kudritzki, 1988, 1998, 2002
and references therein for more details). 
The quantities $k, \alpha , \delta$ are known as the force multiplier 
parameters.

The expected dependence of radiative wind efficiency   
on metallicity (in the range from solar to 0.01 solar) 
has been investigated by e.g. Abbott (1982b), Kudritzki et al. (1987),
Leitherer et al. (1992), Kudritzki \& Puls (2000, and references therein), 
Vink et al. 2001. 
The mass-loss rates are predicted to decrease at lower 
metallicities as a power law following
$(Z/Z_{\odot})^{b}$, with $b = 0.5 - 0.8$.

Kudritzki (2002) in his investigation of winds, ionising fluxes and
spectra of very massive stars at very low metallicities has recently extended
the analysis down to 0.0001 solar.
These calculations introduce a new treatment of the radiative line force
and are based on 
revised and updated atomic physics and
line lists  (see e.g. Puls et al. 2000 and references therein).
Following a more realistic consideration of the line
strength distribution function, the usual assumption of constant
force-multiplier parameters ($k$, $\alpha$, $\delta$)  
throughout the atmosphere is replaced 
by the introduction of an explicit dependence of these 
parameters on optical depth and density. This improved treatment of the line
force is needed at extremely low metallicity, because the depth dependence
of the radiative acceleration differs significantly from the case of higher
metallicity. As a consequence, a new
elaborate algorithm to calculate the conditions at the critical point of the
wind is also needed.

%\begin{table}
%\caption{Coefficients of the polynomials (see Eq.~(\ref{eq_3})) for the
%radiative mass-loss formula by Kudritzki (2002)}
%\label{tab_coef}
%\begin{tabular}{lcrrr}
%\noalign{\smallskip}
%\hline
%\noalign{\smallskip}
%$f$ & $T_{\rm eff}$ & $a_0$ & $a_1$ & $a_2$ \\
%\noalign{\smallskip}
%\hline
%\noalign{\smallskip}
%$[Z]_{\rm min}$ & 60000 & -3.40 & -0.40 & -0.65 \\
%                & 50000 & -3.85 & -0.05 & -0.60 \\
%                & 40000 & -4.45 &  0.35 &  -0.80 \\
%$Q_{\rm min}$   & 60000 & -8.00 & -1.20 &  2.15 \\
%                & 50000 & -10.35 & 3.25 & 0.00 \\
%                & 40000 & -11.75 & 3.65 & 0.00 \\
%$Q_0$           & 60000 & -5.99 & 1.00 & 1.50 \\
%                & 50000 & -4.85 & 0.50 & 1.00 \\
%                & 40000 & -5.20 & 0.93 & 0.85 \\
%\noalign{\smallskip}
%\hline
%\noalign{\smallskip}
%\end{tabular}
%\end{table}

The results of Kudritzki's stellar wind calculations are fitted
by means of analytic formulas, that give
the mass-loss rate as a function of effective temperature,
luminosity and metal abundance for very massive stars with masses above
100 $M_{\odot}$:
\begin{equation}
\log \dot M = q_1([Z]-[Z_{\rm min}])^{0.5} + Q_{\rm min}\,\,\,\,\,\,\,\, 
{\rm in} \,\, M_{\odot}\, {\rm yr}^{-1}
\label{eq_1}
\end{equation}
where we define 
\begin{equation}
q_1 = Q_0 -Q_{\rm min} (-[Z]_{\rm min})^{-0.5}\,,\,\,\, 
[Z] = \log(Z/Z_{\odot})\, 
\label{eq_2}
\end{equation}
with $Z$ being the metallicity (abundance in mass fraction).
The quantities $[Z]_{\rm min}$, $Q_{\rm min}$, and $ Q_0$
are expressed as polynomials of the luminosity, in the form
\begin{equation}
f = a_0 + a_1 \L + a_2 \L^2\,\,\,\,\,\,\,\, {\rm with}\,\,\,
\L = \log (L/L_{\odot})-6.0.
\label{eq_3}
\end{equation}
The coefficients $a_i$ are given as a function of the 
effective temperature (see table 3 in Kudritki 2002), 
in the range from 40000 to 60000 K.
%(see Table \ref{tab_coef}).

The above formalism is applied with the following assumptions:
\begin{itemize}
\item 
At any time the metallicity -- that enters Eqs.~(\ref{eq_1})-(\ref{eq_2}) -- 
is set $Z=max(Z_{\rm min}, Z_{\rm eff})$. 
This means that when evaluating the radiative
mass-loss rate we do not assume a truly zero metal content, 
but the maximum between the lowest value in the validity range
of Kudritzki's formula ($Z_{\rm min} = 10^{-4}\times Z_{\odot}$), 
and $Z_{\rm eff}=1-X-Y$.
This latter denotes 
the effective current metallicity, with $X$ and $Y$  
being the  hydrogen and helium abundances, respectively.
In this way we may account, to a first approximation, for 
any possible surface enrichment in metals (mainly CNO elements, such
that $Z_{\rm eff} > Z_{\rm min}$) exposed
to the surface by stellar winds or convective  mixing. 
\item For $T_{\rm eff} > 60000$ K, we choose not to extrapolate the
fit formula beyond its validity domain, but constrain
$T_{\rm eff}=min(T_{\rm eff}, 60000)$.
It should be noticed that most of the main-sequence lifetime
of VMOs is spent at $T_{\rm eff}> 60000$.
\end{itemize}

\subsection{Effect of rotation}
\label{ssec_rot}
As shown by e.g. Friend \& Abbott (1986) and Pauldrach et al. (1986), 
radiative mass-loss rates 
are enhanced by rotation by a factor that depends on the ratio, 
$v_{\rm rot}/v_{\rm crit}$, between the rotational velocity 
(usually taken at the equator),
and the critical or break-up velocity.
This latter is defined by the vanishing of the
net radial acceleration contributed by the gravitational, radiative, 
and centrifugal forces.

In our work we apply the analytic recipe presented in Maeder \& Meynet (2000,
and earlier works of the series referenced therein), where the 
results of a detailed quantitative analysis of stellar rotation
are conveniently synthesised.
According to Maeder \& Meynet's scheme, the mass-loss rate
from a rotating  star can be expressed  as a function
of the purely radiative $\dot M_{\rm rad}$, and the rotational
correction factor:
\beq
	\dot M(v_{\rm rot}) = F_{\Omega} \times \dot M_{\rm rad}(v_{\rm rot}=0)\,,
\label{eq_mdotrot}
\eeq
with the correction factor given by
\beq	
	F_{\Omega} = \frac{(1 - \Gamma)^{\frac{1}{\alpha}-1}}
       {[1 -(T_{\Omega} + \Gamma)]^{\frac{1}{\alpha}-1}}\,,
\label{eq_gof}
\eeq
where  
$\Gamma$ is the Eddington factor for electron screening opacity, 
$\Omega$ is the angular velocity,  $\rho_{\rm m}$ is the internal mean 
density, and 
$\alpha$ is the force multiplier parameter
corresponding to the slope of the line strength
distribution.

The effect of rotation is contained in the term
\beq  
T_{\Omega} = \frac{\Omega^2}{2 \pi G \rho_{m}} \simeq \frac{4}{9}
\frac{v_{\rm rot}^2}{v_{\rm crit}^2}\,,
\eeq
where the 
break-up velocity is evaluated as 
\beq 
	v_{\rm crit} = \Big(\frac{2}{3} \frac{G M }{R}\Big)^{\frac{1}{2}}
\eeq
with clear meaning of the quantities.

We refer to the classical $\Omega$-limit when the centrifugal acceleration
balances the gravitational one, that is when 
$v_{\rm rot}=v_{\rm crit}$ and the effects of radiation  
can be neglected. This is not the case of VMOs, as they are characterised 
by large Eddington factors. 
The so-called $\Omega \Gamma$-limit applies, instead,  
when the effects of both
radiation and rotation concur to balance the gravitational acceleration.
This corresponds mathematically to the 
divergency of the factor $F_{\Omega}$, that is when the denominator
of Eq.~(\ref{eq_gof}) vanishes.

It is worth noticing that for $v_{\rm rot}=v_{\rm crit}$, the
denominator of Eq.~(\ref{eq_gof}) becomes 
$\sim [0.361-\Gamma]^{\frac{1}{\alpha}-1}$ and  
the  $\Omega \Gamma$-limit could be met 
for values $\Gamma \ge 0.639$, that
is for luminosities fainter than the Eddington limit.
Viceversa, if  $\Gamma > 0.639$, the  $\Omega \Gamma$-limit could be
already reached for rotational velocities lower than $v_{\rm crit}$,
that is below the $\Omega$-limit.

If $\Omega \Gamma$-limit is attained, then Eq.~(\ref{eq_mdotrot}) 
becomes meaningless,
and the theory does not indicate  any straightforward simple treatment for 
such critical occurrence. 
For the sake of simplicity, we handle this circumstance by  
assuming a constant value for the mass-loss rate, $\dot M_{\rm crit}$,
its value being a free parameter to be specified
in our calculations.

The evolution of the rotational velocity
$v_{\rm rot} = \Omega(R) \times R$ (where $R$ is the surface radius) 
is derived under simple assumptions, namely:   
\begin{itemize} 
\item Spherical symmetry
\item Rigid-body rotation, $\Omega(r)=\Omega$
\item Conservation of the current total angular momentum 
over any evolutionary time step 
\item The equations of stellar structures are kept unchanged, i.e.
without including any rotational term.
\end{itemize}
In summary, we proceed as follows. 
Let us suppose that at a given time $t$ the total angular
momentum of the star is
\begin{equation}
\mathcal{L}(\Omega) = \Omega \times \mathcal{J} \,, 
\end{equation}
where
\begin{equation}
\mathcal{J}  = \frac{8 \pi}{3}\int_0^R \rho(r)\, r^4 \ud r 
\end{equation}
is the moment of inertia of the stellar sphere.
All quantities, like the stratification in radial coordinate $r$, 
density $\rho$ and surface radius $R$, are evaluated at $t$.

If during the evolutionary time step, $\delta t = t^{\prime} - t$, 
the star has lost a shell of mass $\delta M$, 
comprised between radii $\bar{r}$ and $R$, then the total angular momentum 
at time $t^{\prime}$ is given:
\begin{eqnarray}
\mathcal{L^{\prime}}(\Omega^{\prime}) & = & \Omega^{\prime} \times
\frac{8 \pi}{3}\int_0^{R^{\prime}} \rho^{\prime}(r^{\prime}) {r^{\prime}}^4 
\ud r^{\prime}  \\
\nonumber
  &  = & \mathcal{L}(\Omega) - \Omega \times \frac{8 \pi}{3}\int_{\bar{r}}^R 
\rho(r) \, r^4 \ud r  
\end{eqnarray}
where all primed quantities are now referred to time $t^{\prime}$.

In this way we can account for the decrease of the total angular momentum 
due to mass loss (see Heger \& Langer 1998), as well as 
the effects of possible spinning-up (slowing-down)
of rotation due to contraction (expansion) of the star.

The free rotation parameters to be specified before calculating 
each model are: i) the initial rotational velocity $v_{{\rm rot},0}$, 
and ii), the force multiplier parameter $\alpha$, and 
the critical mass loss rate $\dot M_{\rm crit}$ at the 
$ \Omega\Gamma$-limit.
In our calculations we adopt $v_{{\rm rot},0} = 500 $ km $s^{-1}$;
$\alpha=0.1,\, 0.4$; and
$\dot M_{\rm crit} = 10^{-3}\, M_{\odot}$ yr$^{-1}$.

We set $\alpha$ lower than the standard value of $2/3$ valid
for hydrogenic ions since, according to Kudritzki (2002),  
$\alpha$ is found to decrease at decreasing metallicities. 
We notice that the lower $\alpha$, the larger
the correction factor due to rotation $F_{\Omega}$ (see Eq.~(\ref{eq_gof})).  
However, since our results for the case $\alpha=0.1$ do not substantially
differ from those for $\alpha=0.4$, we will show and discuss  
only these latter.

Finally, we choose to set $\dot M_{\rm crit}$ of the same order of magnitude 
as the very large mass-loss rates measured 
in Luminous Blue Variables (LBV), yellow and blue super-giants 
(see e.g. Leitherer 1997).

\begin{table*}
\caption{Evolutionary properties of zero-metallicity VMOs
for the adopted mass-loss prescriptions. 
From left to right the table entries read: the stellar initial mass; 
the H- and He-lifetimes; 
the mass of ejecta, carbon and helium cores at central C-ignition;
the wind yields of helium, and C,N,O elements.
All masses are
expressed in $M_{\odot}$, including the chemical yields. 
The nuclear lifetimes are given in yr.}
\label{tab_mod}
\begin{tabular}{rccrrrccccc}
\noalign{\smallskip}
\hline
\noalign{\smallskip}
\multicolumn{1}{c}{$M$} &
\multicolumn{1}{c}{$\tau_{\rm H}$} &
\multicolumn{1}{c}{$\tau_{\rm He}$} &
\multicolumn{1}{c}{$\Delta M_{\rm ej}$} &
\multicolumn{1}{c}{$M_{\rm C}$} &
\multicolumn{1}{c}{$M_{\rm He}$} &
\multicolumn{1}{c}{y($^{4}$He)} &
\multicolumn{1}{c}{y($^{12}$C)} &
%\multicolumn{1}{c}{y($^{13}$C)} &
\multicolumn{1}{c}{y($^{14}$N)} &
\multicolumn{1}{c}{y($^{16}$O)} \\
\noalign{\smallskip}
\hline
\noalign{\smallskip}
\multicolumn{11}{l}{\underline{MODELS with $v_{\rm rot} = 0$; 
$\dot M =\dot M_{\rm rad}$}} \\
\noalign{\smallskip}   
120 &  2.745E+06  &   2.557E+05   &    0.02    &   50.94    &     59.95    &     4.090E-03   &      6.805E-18     &    5.536E-16     &    1.056E-17 &  \\
250  &   2.219E+06  &   2.281E+05    &      0.19    &    117.29   &     132.93    &     8.655E-07   &      2.956E-16     &    1.966E-14   &      3.595E-16 & \\
500   &   1.943E+06   &  2.132E+05     &    11.38    &    243.00    &    263.89   &      1.725E-02    &     5.251E-12   &      4.870E-10     &    8.745E-12 & \\
750  &    1.852E+06   &  2.111E+05   &     146.20    &    343.84    &    390.89    &     1.324E+01    &     4.229E-09    &     3.983E-07    &     4.842E-09 & \\
1000  &    1.861E+06   &  2.153E+05    &    581.29    &    346.17    &    418.67     &    1.733E+02    &    4.109E-01    &     2.057E-05     &    3.089E-01 & \\
\noalign{\smallskip}
\hline
\noalign{\smallskip}
\multicolumn{11}{l}{\underline{MODELS with $v_{\rm rot,0} = 500$ km s$^{-1}$}} \\ 
\noalign{\smallskip}
120 &  2.745E+06 &  2.550E+05   &     1.43    &   50.92   &    59.99   &    1.587E-08    &   3.039E-18  &     2.478E-16  &     4.728E-18 & \\
250 &  2.219E+06  & 2.283E+05    &    3.35   &   117.50    &  132.70  &     3.853E-06    &   1.085E-15  &     8.520E-14    &   1.587E-15 & \\
500 &  1.941E+06 &  2.169E+05   &    20.67    &  245.22   &    268.93  &      5.848E-01   &     1.767E-10   &    1.507E-08    &   2.877E-10 & \\
750 &  1.853E+06  & 2.092E+05   &   158.30  &    343.63   &   385.81  &     1.612E+01    &   5.002E-09   &    4.682E-07    &   5.742E-09 & \\
1000 &  1.862E+06 &  2.155E+05    &  581.41   &   344.27   &   414.40   &    1.697E+02   &    1.184E+00    &   1.733E-05    &   1.501E+00 & \\
\noalign{\smallskip}
\hline
\noalign{\smallskip}
\end{tabular}
\end{table*}
%%%%%%%%%%%%%%%%%% FIGURE %%%%%%%%%%%%%%%%%%%%%%%%%%%%%%%%%%%%%%%%%%%%%%%
\begin{figure}
\resizebox{\hsize}{!}{\includegraphics{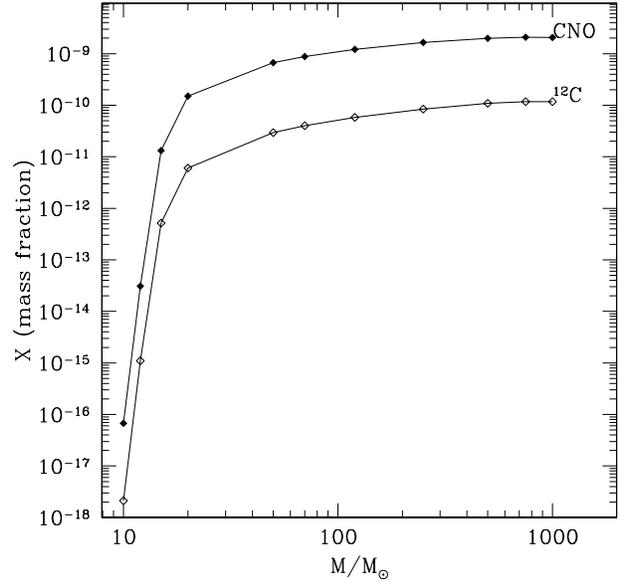}}
\caption{CNO and $^{12}$C central abundances (in mass fraction) 
when $5 \%$ of hydrogen
has been burnt in the convective core, as a function of the initial
stellar mass }
\label{fig_cno}
\end{figure}
%%%%%%%%%%%%%%%%%%% FIGURE %%%%%%%%%%%%%%%%%%%%%%%%%%%%%%%%%%%%%%%%%%%

\subsection{General characteristics}
\label{ssec_general}
Let us now first consider the general features that
characterise the evolution of zero-metallicity VMOs, 
and then move to discuss the efficiency of mass loss.

It is well known that in stars originally lacking in CNO elements
the (pre-)main sequence gravitational contraction
of the central regions may proceed to such an extent that 
the density and temperature conditions allow 
the ignition of the triple$-\alpha$ reaction, with consequent activation of   
the CNO-cycle. 
In stars with $M\, \la\, 20\, M_{\odot}$, 
the first occurrence of the triple$-\alpha$ reaction 
usually takes place
after core H-burning has already started via the p-p reactions, the time delay
between the two events reducing at increasing stellar mass.
Eventually, in more massive stars, with $M\, \ga\, 20\, M_{\odot}$, 
the first synthesis of primary $^{12}$C is practically simultaneous 
with the onset of core H-burning,  
which proceeds via the CNO cycle since the very initial stages.

This is illustrated in
Fig.~\ref{fig_cno}, showing the
primary CNO and $^{12}$C abundances in the convective
core at the beginning of H-burning, as a function
of the initial stellar mass.
It is worth noticing that $X({\rm CNO}) \sim X(^{14}$N),
because as soon as the triple-$\alpha$ reaction produces 
some $^{12}$C, most of it is immediately burnt into  $^{14}$N,
given the high temperatures at which the CNO-cycle operates. 
For  $M\, \ga\, 20\, M_{\odot}$,
the total $X$(CNO) has already grown to  $\sim 10^{-10} - 10^{-9}$, 
which guarantees  the
primary role of the CNO-cycle as nuclear energy source for the star
during the whole main sequence phase (see Paper I).
The abrupt drop of $X$(CNO) at lower masses indicates that 
the full activation of the CNO-cycle is delayed to later stages
of H-burning, initially proceeding via the p-p reactions
(see also figure 5 in Paper I).

Various model predictions for relevant quantities 
are presented in Table~\ref{tab_mod}.
Let us first consider the nuclear lifetimes of core H- and He-burning 
($\tau_{\rm H}$ and $\tau_{\rm He}$)  as a function of the initial 
stellar mass.
In both cases the trend is initially decreasing  
at larger masses, until it flattens out towards a nearly constant
value of roughly $2 \times 10^6$ yr  for $\tau_{\rm H}$ 
and $2 \times 10^5$ yr for $\tau_{\rm He}$, towards the most massive
models ($M \ga 500 M_{\odot}$). 
It is worth briefly recalling the reason for this, 
with the aid of approximate scaling relations 
(see Kippenhahn \& Weigert 1994),
e.g. the definition of the nuclear H-burning lifetime, 
$\tau_{\rm H} \propto M/L$, 
and the mass-luminosity relation, $L \propto M^{\eta}$, so that
$\tau_{\rm H} \propto  M^{1-\eta}$.
As the exponent $\eta$ is typically 3-3.5 (say in the range from about 1 
to 100 $M_{\odot}$), the nuclear lifetime is expected to decrease 
at increasing stellar mass. The almost invariance of $\tau_{\rm H}$   
seen at very large masses is the result of the growing contribution of
radiation to the total pressure, which makes the mass-luminosity relation 
approach a linear proportionality  ($\eta \sim 1$ and $L \propto M$), 
so that $\tau_{\rm H} \approx const$.

%%%%%%%%%%%%%%%%%% FIGURE %%%%%%%%%%%%%%%%%%%%%%%%%%%%%%%%%%%%%%%%%%%%%%%
\begin{figure*}
\begin{minipage}{0.70\textwidth}
\resizebox{\hsize}{!}{\includegraphics{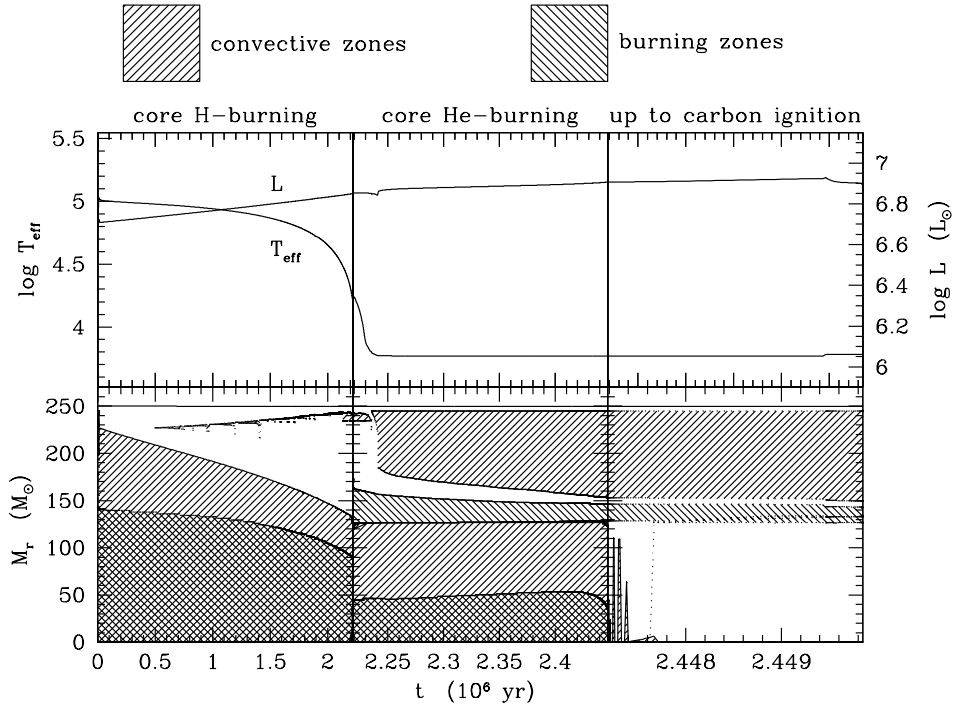}}
\end{minipage}
\hfill
\begin{minipage}{0.28\textwidth}
\caption{Evolutionary properties of the 250 $M_{\odot}$ model with 
the prescription for the purely radiative mass loss.
Top panel: Stellar luminosity and effective temperature as a function
of time during the major nuclear burnings up to central carbon ignition. 
Bottom panel: Convective and burning regions (in mass coordinate 
from the centre to the surface) across the stellar 
structure. The upper solid line represents the mass coordinate 
of the stellar surface}
\label{fig_250kud}
\end{minipage}
\end{figure*}
%%%%%%%%%%%%%%%%%%% FIGURE %%%%%%%%%%%%%%%%%%%%%%%%%%%%%%%%%%%%%%%%%%%
%%%%%%%%%%%%%%%%%% FIGURE %%%%%%%%%%%%%%%%%%%%%%%%%%%%%%%%%%%%%%%%%%%%%%%
\begin{figure*}
\begin{minipage}{0.70\textwidth}
\resizebox{\hsize}{!}{\includegraphics{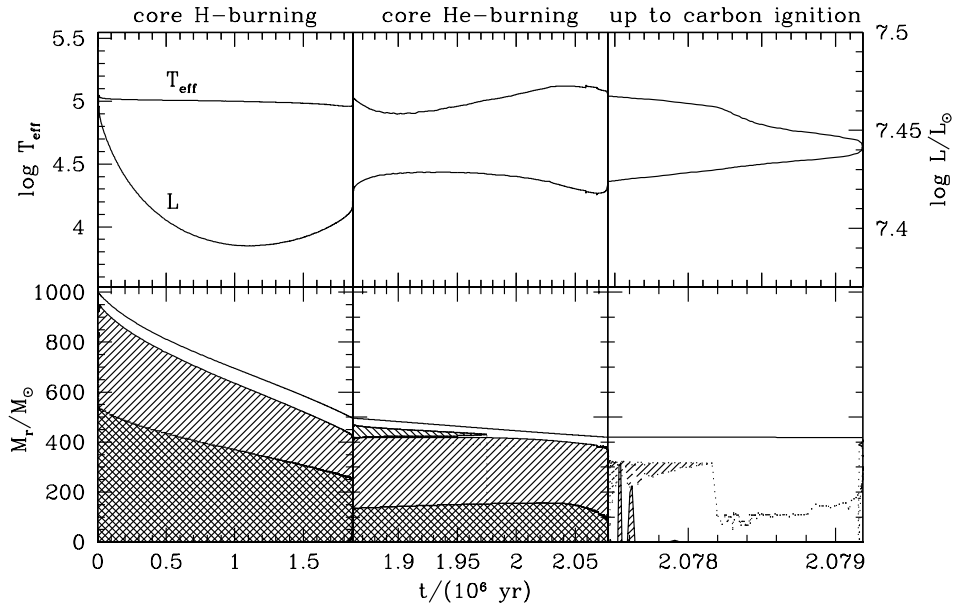}}
\end{minipage}
\hfill
\begin{minipage}{0.28\textwidth}
\caption{The same as in Fig.\protect\ref{fig_250kud}, but referring
to the evolution of the 1000 $M_{\odot}$ model calculated with
the prescription for radiative-rotational mass-loss. The assumed
initial rotational velocity is $v_{\rm rot} = 500$ km $s^{-1}$.
An almost identical plot is obtained for the model with the same initial
mass, but assuming the purely radiative mass-loss rates}
\label{fig_1000vrot5}
\end{minipage}
\end{figure*}
%%%%%%%%%%%%%%%%%%% FIGURE %%%%%%%%%%%%%%%%%%%%%%%%%%%%%%%%%%%%%%%%%%%

One common feature of the VMO models is the 
development of very large convective cores during both 
the H- and He-burning phase, 
with a typical fractional extension amounting to 70 -- 90 $\%$ 
of the total mass.
Moreover, at the exhaustion of hydrogen in the core central convection   
does not disappear, and the onset of core He-burning
immediately follows because of the already high temperatures attained
($\log T \sim 8.2-8.3$). This implies that the
intermediate phase of H-shell burning is not expected for these stars.  

%%%%%%%%%%%%%%%%%% FIGURE %%%%%%%%%%%%%%%%%%%%%%%%%%%%%%%%%%%%%%%%%%%%%%%
\begin{figure}
\resizebox{\hsize}{!}{\includegraphics{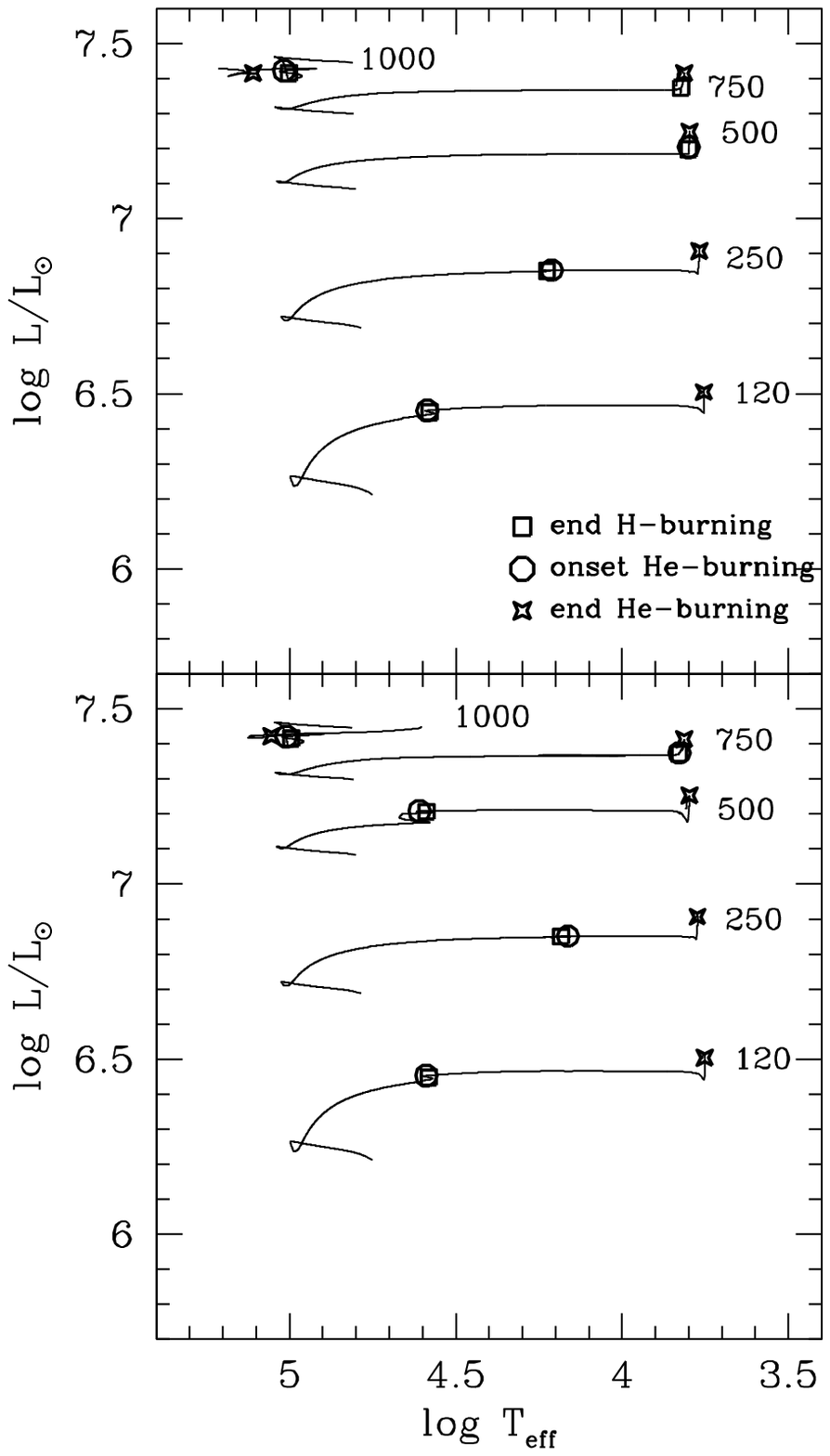}}
\caption{Zero-metallicity stellar tracks of VMO with initial masses 
of 120, 250, 500, 750, 1000 $M_{\odot}$.
The evolution covers the H- and He-burning phases
up to ignition of central carbon.
{\bf Top panel:} Purely radiative mass-loss prescription;
{\bf Bottom panel:} Rotational-radiative mass-loss prescription}
\label{fig_hr}
\end{figure}

Two examples are illustrated 
in Figs.~\ref{fig_250kud} and \ref{fig_1000vrot5}, 
that show the evolution of both convective and burning regions in 
two  models with initial masses of $250\, M_{\odot}$ and $1000\, M_{\odot}$,
for two choices of the mass-loss prescription.
From these plots we can already see that 
new products synthesised by nuclear burnings 
may be exposed to the stellar surface 
because of i) the inward penetration of the convective envelope across
a chemical profile left by the gradual recession of the 
convective core, and/or ii) the progressive peeling of the stars
due to mass loss. This point will be discussed in more detail
in the following sections.

Figure \ref{fig_hr} displays the evolutionary tracks
in the H-R diagram for the two sets of VMOs, calculated adopting the
prescriptions for purely radiative  
and rotational-radiative mass loss, 
respectively.
In all cases  central H-burning is mostly spent at 
high effective temperatures,
regardless the efficiency of mass loss. 
We can notice that the locus of points describing the 
onset of core He-burning (practically coincident with
the end of core H-burning) bends  towards lower $T_{\rm eff}$  
at increasing stellar mass.
On the other hand, the He-burning
phase takes place  in different regions of the H-R diagram, depending 
on stellar mass and mass loss.
Soon after the onset of He-burning, models  
with $120 M_{\odot} \le M \le 750$ are able to reach their Hayashi line,
where they remain up to central carbon ignition.
The evolution of models with $M=1000\, M_{\odot}$  
takes place entirely  at high effective temperatures, essentially due 
significant mass loss.
Next section will examine this aspect in mode detail.

Finally, Fig.~\ref{fig_mcore} shows the expected  behaviour of the
mass of the He- and CO- core at the stage of central carbon ignition,
as a function of the initial mass in the domains of massive and very
massive stars ($ 8 \, M_{\odot} \le M \le 1000\, M_{\odot}$).
The relation is almost linear, without any important effect due to
mass loss. Our predictions indicate that the interval 
$M_{\rm He} = 64$ to $ 133\, M_{\odot}$, which would produce  
pair-instability supernovae according to Heger \& Woosley (2002)'s 
recent calculations, correspond to stellar progenitors
with initial masses  $M = 127$ to $252\, M_{\odot}$
(see also El Eid et al. 1983, 
Ober et al. 1983, Bond et al. 1984, Carr et al. 1984). 

\subsection{Mass-loss efficiency} 
\label{ssec_mlosshr}
We will discuss now the efficiency of mass loss in relation to the
two driving mechanisms here considered, namely: radiation 
(Sect.~\ref{ssec_rad}) and rotation (Sect.~\ref{ssec_rot}). 
Stellar ejecta are presented in Table \ref{tab_mod}.

\paragraph{Models with $\dot M = \dot M_{\rm rad}$.}
In general, the purely-radiative mass-loss rates remain quite low 
during the whole evolution of VMO models with  $M < 500 \, M_{\odot}$. 
For instance, the 120 $M_{\odot}$ model returns  only 0.02 $M_{\odot}$
to the interstellar medium.
In this mass range, 
the general properties are practically the same ones as those
predicted for constant-mass evolution.
As reasonable extrapolation we could expect that at  lower masses
($< 120\, M_{\odot}$, hence lower luminosities), 
the inefficiency of mass loss should  even more
marked, which supports the scenario of constant-mass evolution
of Pop-III stars discussed in Paper I. 
Another  common characteristic of the models
 with $M = 120  - 500 \, M_{\odot}$
is the morphology of the tracks in the H-R diagram, and
hence the location of major nuclear burnings.
Actually, if the H-burning phase is spent in regions of high
effective temperature (typically at $\log T_{\rm eff} \sim 4.8-5.0$),
the subsequent He-burning phase takes place far away from there,  
as the star evolves along its Hayashi line (typically at 
$\log T_{\rm eff} \sim 3.7-3.8$).
Indeed, the tracks of models with   
$M < 500 \, M_{\odot}$ are characterised by quite a large excursion
in $T_{\rm eff}$, that is performed at almost constant luminosity.

Similar considerations on the morphology of the tracks in the H-R diagram 
apply to the $750\, M_{\odot}$ model as well, though
in this case the efficiency of mass loss is somewhat more
significant, the total ejecta amounting to about $20 \, \%$ of the
initial mass (see Table~\ref{tab_mod}).

A different behaviour characterises the most massive model here considered,
with $M =1000\, M_{\odot}$. 
In this case,
radiative mass loss is predicted to be strong 
since the very beginning of the MS phase, due to the combination
of large luminosities and effective temperatures.
The entire evolution is confined in regions of the H-R diagram 
of high effective temperatures, and    
the corresponding stellar ejecta up to the central carbon ignition 
are considerable, reaching to more than half the initial mass.

The net result is that radiation-driven stellar winds
do not produce appreciable mass loss from massive zero-metallicity
stars, except for the most massive models,  with $M > 750 \, M_{\odot}$. 
Our finding  supports previous indications  in this sense by 
Kudritzki (2000) and Bromm et al. (2001). 

\begin{figure}
\resizebox{\hsize}{!}{\includegraphics{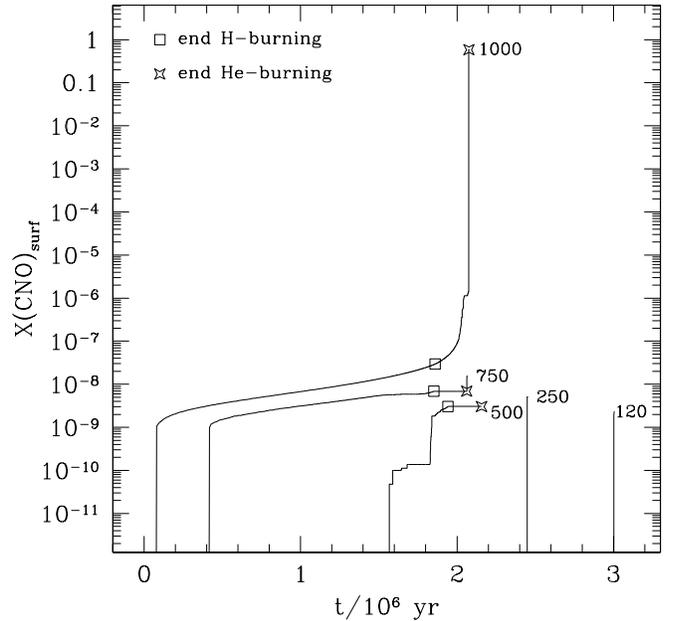}}
\caption{Surface abundance of CNO elements (in mass fraction)
as a function of the evolutionary time, from the ZAMS up to carbon ignition
in the core. Curves correspond to different initial masses (as indicated)
of VMOs with initial zero metallicity. The radiative mass-loss prescription
is applied. Note that in most cases $X$(CNO)$_{\rm surf}$ remains 
lower than $Z_{\rm min}$ assumed in the radiative mass-loss formula
(see Sect.~\protect{\ref{ssec_rad}}) }
\label{fig_surfchem}
\end{figure}

It is worth making a final remark by looking at Fig.~\ref{fig_surfchem}, 
which displays the degree of chemical self-pollution at the surface, 
in terms of CNO abundance, produced by the combined effect of mass loss
and convective mixing. As we see in most models the maximum enrichment
does not exceed $10^{-9}-10^{-8}$, that is the critical CNO abundance
required in stars with original metal-free composition to fully activate
the CNO-cycle as dominant energy source (see Sect.~\ref{ssec_general} 
and Fig.~\ref{fig_cno}). This occurs when the mass-loss front 
is only able to penetrate into the chemical profile generated by the recession
of the convective core during the MS phase, as it is the case of the 
500 and 750 $M_{\odot}$ models.  
Models with lower masses, 120 and 250  $M_{\odot}$, show comparable
self-enrichment, that is however reached after the end of the He-burning phase,
not as a consequence of mass loss but because of deep convective dredge-up
while on the Hayashi line.
It follows that 
such degree of surface chemical enrichment should not produce 
enough resonant metal lines in the atmosphere to activate
efficient radiation-driven mass-loss.  

%%%%%%%%%%%%%%%%%%% FIGURE %%%%%%%%%%%%%%%%%%%%%%%%%%%%%%%%%%%%%%%%%%%
%%%%%%%%%%%%%%%%%%% FIGURE %%%%%%%%%%%%%%%%%%%%%%%%%%%%%%%%%%%%%%%%%%%

On the contrary, the most massive 1000 $M_{\odot}$ model exhibits a remarkable
increase of the CNO abundance during the He-burning phase, due to the
progressive reduction of the stellar mass.
As shown in Fig.~\ref{fig_1000vrot5} the  H-burning shell is 
extinguished and eventually, close to the exhaustion of central helium, 
the mass-loss front penetrates into regions previously reached 
by convective core. Then, significant amounts of carbon and oxygen are
exposed at the surface ($X$(CNO)$\approx 0.6$).
Whenever this circumstance may trigger substantial radiative mass-loss
is an interesting  point to be considered, and it would deserve a dedicated 
study in the framework of the radiation-driven-winds theory. 
Finally,  it is worth remarking that an additional chemical pollution 
of the stellar surface may be caused by rotationally-induced mixing
(e.g. Heger \& Langer 2000; Meynet \& Maeder 2002), 
which is not taken into account in the present analysis.

\paragraph{Models with $\dot M = \dot M(v_{\rm rot})$.}
The amplification effect  on mass loss due to stellar rotation 
turns out rather  limited. 
As we can see, the evolutionary tracks in bottom panel of 
Fig.~\ref{fig_hr} are almost indistinguishable from those 
in the top panel.
Actually, the total ejecta are larger than those for models with radiative 
rates, but not such to drastically alter the evolutionary properties
just presented. 

%%%%%%%%%%%%%%%%%% FIGURE %%%%%%%%%%%%%%%%%%%%%%%%%%%%%%%%%%%%%%%%%%%%%%%
\begin{figure}
\resizebox{\hsize}{!}{\includegraphics{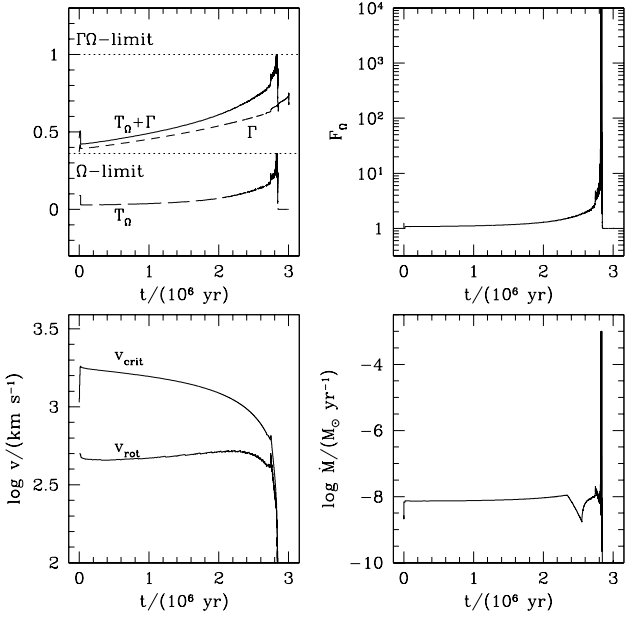}}
\caption{Evolution of rotational parameters 
referring to the 120 $M_{\odot}$ model, calculated 
with $V_{\rm rot,0}=500$ km s$^{-1}$, and $M_{\rm crit} = 10^{-3}\,\,
M_{\odot}$ yr$^{-1}$. 
Top-left panel: Relevant terms determining
the rotation correction factor given by Eq.~(\ref{eq_gof}).
The $\Omega$- and $\Omega\Gamma $-limits are indicated  
by horizontal lines delimiting the  maximum values 
for $T_{\Omega}$ and $T_{\Omega}+\Gamma$, respectively.
Top-right panel: Rotation correction factor.
Bottom-left panel: Evolution of surface ($v_{\rm rot}$) and critical 
($v_{\rm crit}$) rotation velocities.
Bottom-right panel: mass-loss rate}
\label{fig_rot120}
\end{figure}
%%%%%%%%%%%%%%%%%%% FIGURE %%%%%%%%%%%%%%%%%%%%%%%%%%%%%%%%%%%%%%%%%%%
%%%%%%%%%%%%%%%%%% FIGURE %%%%%%%%%%%%%%%%%%%%%%%%%%%%%%%%%%%%%%%%%%%%%%%
\begin{figure}
\resizebox{\hsize}{!}{\includegraphics{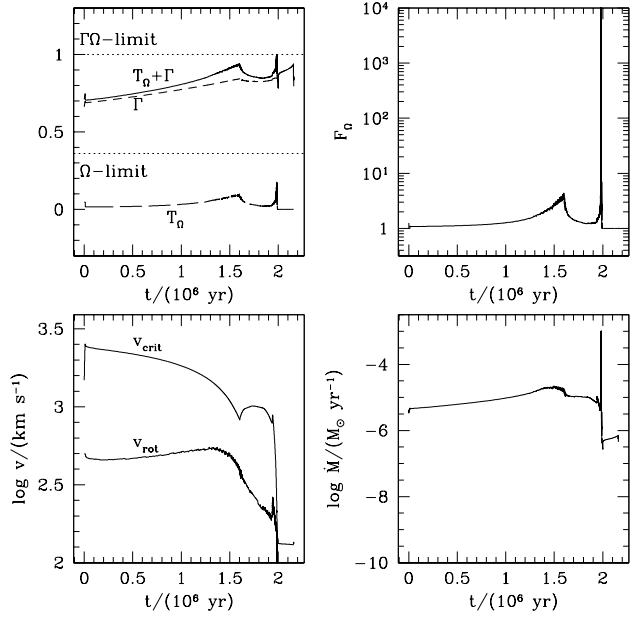}}
\caption{The same as in Fig.~\protect\ref{fig_rot120}, but for the 500
$M_{\odot}$ model}
\label{fig_rot500}
\end{figure}
%%%%%%%%%%%%%%%%%%% FIGURE %%%%%%%%%%%%%%%%%%%%%%%%%%%%%%%%%%%%%%%%%%%
 
To better understand the results we refer to Figs.~\ref{fig_rot120} and
\ref{fig_rot500}, showing the evolution of rotational parameters and 
mass-loss rates. 
In both cases the correction factor $F_{\Omega}$
(Eqs.~(\ref{eq_mdotrot}) and (\ref{eq_gof})) keeps close to one during 
nearly the whole MS phase, so that the mass-loss rate remains quite low
($\sim 10^{-8}\, M_{\odot}$ yr$^{-1}$).

Eventually, during the stages just preceding and following the H-consumption 
in the core, the 120 $M_{\odot}$ model reaches 
both the  $\Omega$- and $ \Omega \Gamma$-limits
at the same time.
This result is supported by recent analyses of rotating 
stars at low metallicities. In fact, as discussed by Meynet \& Maeder
(2002) the  approach of the  $\Omega\Gamma$-limit 
is just favoured in massive and luminous stars at 
very low- or even zero metallicity. 
This circumstance should be the consequence of two concurring factors, namely: 
i) the nearly total conservation of the initial angular momentum 
during the almost constant-mass evolutionary stages, that would  
lead to the natural settling 
of the break-up rotation, hence the $\Omega$-limit; and
ii) the large values of the Eddington factor in massive and luminous
stars. In practice, both factors favour the divergency 
of the correction factor given by Eq.~(\ref{eq_gof}).

With respect to the former point i), we recall that various 
analyses (e.g. Heger \& Langer 1998, 2000; Meyner \& Maeder 2000,
Maeder \& Meynet 2001) have shown that a massive star evolving 
at (nearly) constant mass could approach or even reach the break-up rotation, 
towards the  end of the MS phase (or in a blue-loop 
off the Hayashi line). In fact during these stages, 
the progressive stellar contraction would result in increasing 
surface rotational velocities (the so-called ``spin-up phases'').
This is exactly what we find in our 120 $M_{\odot}$ model displayed in 
Fig.~\ref{fig_rot120}.  
At this stage the stellar mass is practically the same
($\sim 119.98\,M_{\odot} $) as the initial one.

However this critical regime, 
that we describe by assuming an extremely high mass-loss rate $M_{\rm crit}$,
is maintained only for very short. In fact,
the reduction of  total angular momentum
carried away by the outermost ejected layers, and 
the increase of the radius due to the 
subsequent expansion (following the beginning of core He-burning)
make rotation to slow down (see bottom-left panel of Fig.~\ref{fig_rot120}), 
with consequent departure from the  $\Omega$ and $ \Omega \Gamma$-limits.

Then, the star evolves rapidly towards lower effective temperatures
onto its Hayashi line, which causes a further drastic decrease of the 
rotational velocity. 
The total ejecta with the inclusion of the rotational effect
is almost negligible, $~\sim 1.5 M_{\odot}$ (see Table \ref{tab_mod}).

A similar behaviour is found for the $500\, M_{\odot}$ model.
However, in this case the larger luminosity concurs to lead the star 
closer to the $\Omega\Gamma$-limit already during the MS phase.
Then, the radiative mass-loss rate is enhanced by a few times
(see top-right panel of Fig.~\ref{fig_rot500}).
Again, the $\Omega\Gamma $-limit (but not the $\Omega$-limit) 
is briefly touched  during the spin-up phase towards 
the end of the MS phase, 
but soon after abandoned because of the loss of angular momentum
and the expansion of the star.
Compared to the model with the same initial mass but with
radiative mass loss, the net effect of rotation is to increase 
the total ejecta by roughly a factor of two, which still remains very small 
(see Table~\ref{tab_mod}).

At larger masses ($M=750,\, 1000\, M_{\odot}$) the effect of rotation 
turns out even weaker. In fact, 
for these models the radiation-driven mass loss becomes efficient,
which leads to the removal of a significant part of the original angular
momentum already during the MS, 
hence preventing the approaching  of the critical $\Omega$
and $\Omega-\Gamma$-limits. 

\subsection{Cautionary remarks}
\label{ssec_cautr}
At this point 
it is worth making a few cautionary remarks about
the assumptions and simplifications adopted in our work 
to describe the effect of mass loss in primordial massive stars.

As far as the radiative mass-loss is concerned 
(Sect.~\ref{ssec_rad}), it should be pointed out that
the position $Z=max(Z_{\rm min}, Z_{\rm eff})$ and the
treatment for $T_{\rm eff} > 60000$ K may lead to over-estimate
the actual mass-loss rates, since for instance the minimum metallicity,
$Z_{\rm min} \sim 10^{-6}$,  is typically two orders 
of magnitude larger than $Z_{\rm eff}$, (except for the 1000 $M_{\odot}$ 
model).
Moreover, the application of the formula for $\dot M_{\rm rad}$ 
to models with $M > 500 M_{\odot}$ requires to  extrapolate the
predictions above the upper-limit in luminosity 
(i.e. $\logL = 7.03$) of the validity domain.
  
The simplifying assumption
at this stage, without calculations 
for $T_{\rm eff} > 60000$ K and $\logL > 7.03$ available, 
is that the line force parametrisation at higher temperatures 
and luminosities is 
still reasonably approximated
in this way. New calculations extending the temperature and luminosity
domains are certainly needed for the future.

As far as the rotational mass-loss is concerned (Sect.~\ref{ssec_rot}),
we recall that the results are obtained in the ambit
of a very simplified description of stellar rotation
(mainly due to the assumption of solid-body rotation).
It is clear that a strictly correct procedure should consider 
important processes like meridional circulation, shears, and horizontal 
turbulence, which would also 
determine a feed-back on stellar structure
(see e.g. Kippenhahn \& Thomas 1970; Meynet \& Maeder 1997).
Among the main related effects, it turns out that 
the outward transportation of angular momentum 
from the central regions by meridional circulation would 
produce a  differential rotation.
These aspects are not included in the present work.

Rather, our assumption of solid-body rotation would ideally correspond 
to the case of maximum efficiency of convection and shears
in the transport of angular momentum 
from the inner to the outer regions of the star. In this way even a 
relatively small loss of material from the surface leads to a substantial
loss of the total angular momentum, and consequently to the 
slow-down of  rotation. 

On the other hand, in the context of a detailed treatment of  
rotation -- coupled to stellar structure and including also
the role of meridional circulation --  we may expect that the outward
transport of  angular momentum is less efficient.
In fact, detailed computations accounting for differential rotation 
(e.g. Meynet \& Maeder 2000, 2002) indicate that i) 
the angular velocity increases progressively 
with time towards the centre,  and ii) the break-up rotation 
could be reached earlier and maintained for longer periods.
Moreover, the possible effect of anisotropic mass loss
(Maeder 2002) could further reduce the loss of angular momentum and lead
to earlier break-up.
Finally, two additional effects are expected to gain increasing 
importance at lower metallicity, namely:
i) faster core rotation and more efficient rotational mixing
(Meynet \& Maeder 2002), and ii) possibly larger initial rotational 
velocities (Maeder et al. 1999).

%Correspondingly,  
%the loss of angular momentum from the surface by stellar winds 
%and related spinning down of rotation could be less 
%pronounced compared to the case of rigid rotation.
These considerations lead to the possibility 
that the combined effects of rotation and mass loss may be 	
somewhat under-estimated in our calculations.

It follows that the inclusion of all these aspects might produce 
different results compared to those
presented in this paper. However, since up to now there is 
no dedicated analysis addressing such issues specifically for  
Pop-III stars,
we believe that our simplified exploration of the effect of rotation
on primordial stars should keep its own validity.

\section{The stellar yields}
\label{sec_yields}
If ever formed, massive primordial stars marked the first
step in the chemical evolution of the Universe
(e.g. Oh et al. 2001; Scannapieco et al. 2002).
The efficiency of their contribution to the chemical
enrichment of the pristine gas depends essentially on two
factors, namely: the individual stellar yields,
and the details of the IMF. 

As for the former ingredients, 
predictions of elemental yields 
are available in Woosley \& Weaver (1995) 
for supernovae II explosions produced by zero-metallicity stars 
with initial masses 
in the  range $11 - 40\, M_{\odot}$, and  
in Heger \& Woosley (2002) for pair-instability supernovae,
involving helium core masses in the range 
$64 - 133 \, M_{\odot}$ (corresponding to initial stellar masses
of $140-260 \, M_{\odot}$ under the hypothesis of constant-mass 
pre-supernova evolution).
According to the latter work,
at both lower and larger masses, no explosive nucleosynthesis should 
be injected  into the primordial interstellar medium (ISM) but,  rather, 
trapped into black holes. 
For past works on pair-instability supernovae the reader could refer
to El Eid et al. (1983), Ober et al. (1983), Bond et al. (1984), 
Carr et al. (1984).

In their evolutionary calculations at initial zero metallicity 
Woosley \& Weaver (1995) and Heger \& Woosley (2002) 
did not consider any mass loss during the hydrostatic phases 
(except for that driven by pair-instability pulsation).
Past calculations of the  wind contributions from VMOs  
can be found in the works by El Eid et al. (1982),
Klapp (1983,1984), and more recently 
Portinari et al. (1998). These latter estimates 
are based on the  results of semi-analytic stellar models
developed by Bond et al. (1984).
Our predictions of the wind ejecta from Pop-III VMOs are
given in Table~\ref{tab_mod}   
that presents, for each model with initial mass $M$,
the net wind yields (in $M_{\odot}$)
of $^4$He, $^{12}$C, $^{14}$N, and $^{16}$O, 
the total ejecta $\Delta M_{\rm ej}$, 
and the masses of the He-core ($M_{\rm He}$) and C-core ($M_{\rm C}$) at the ignition of
central carbon.
For the sake of completeness,
the results are grouped on the basis of the adopted mass-loss prescriptions,  
namely: purely-radiative mass loss, and radiative-rotational mass-loss.
However, the differences between the two sets of yields are quite small
(see Sect.~\ref{ssec_mlosshr}). 

It turns out that the wind yields ejected by VMOs
are practically negligible for all masses, except for the 
two most massive models here considered, with $M = 750,\, 1000\, M_{\odot}$.
For instance, the 1000 $M_{\odot}$ model with 
$v_{\rm rot,0} = 500$ km s$^{-1}$ ejects 
about $17\, \%$ of its initial mass as newly synthesized 
helium, and $~ 1-1.5\, M_{\odot}$ in form of carbon 
and oxygen. 
In this particular case (see Fig.~\ref{fig_1000vrot5}), 
during the MS phase the mass-loss
front eats up   
the chemical profile left  by the receding convective core and
the  wind ejecta, returned to the ISM, are mainly enriched 
in $^{4}$He and $^{14}$N. 
Then, during the subsequent He-burning phase, the deep peeling 
keeps going on to such an extent that,
after the ejection of the entire hydrogenic envelope (when the abundance of
central helium is $\sim 0.03$),  
the mass-loss front extends beyond the extinguished H-burning shell,
eventually exposing to the surface the nuclear products  
of partial core He-burning, mainly $^{12}$C and $^{16}$O.
However, the net yields of these elements are not extremely large,
given the short evolutionary time-scales involved from that stage 
up to the central carbon ignition (i.e. the end of calculations).
%%%%%%%%%%%%%%%%%% FIGURE %%%%%%%%%%%%%%%%%%%%%%%%%%%%%%%%%%%%%%%%%%%%%%%
\begin{figure}
\resizebox{\hsize}{!}{\includegraphics{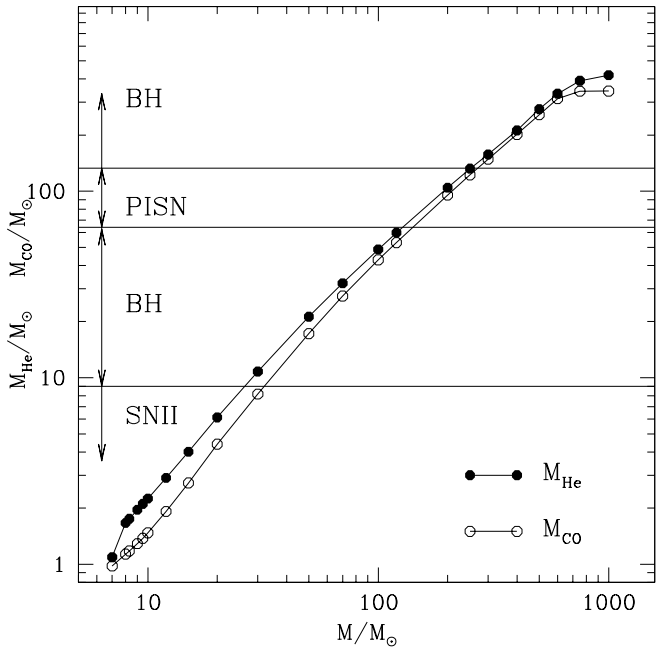}}
\caption{Masses of the He- and CO- cores at the onset of
carbon ignition in the core, as a function of the
stellar initial mass. They correspond to evolutionary calculations
at constant mass for $M \le 100\, M_{\odot}$ (Marigo et al. 2001), 
and  with radiation-driven mass loss for VMOs for 
$120\, M_{\odot} \le M \le 1000, M_{\odot}$.
Relevant ranges of $M_{\rm He}$ are marked and labelled 
-- following  Woosley \& Weaver (1995); Heger \& Woosley (2002) -- 
as a function of the final fate of the stellar progenitor, namely: 
i) black hole (BH) collapse without explosion
($M_{\rm He} > 133 M_{\odot}$);
ii) pair-instability supernovae
(PISN, $64\, M_{\odot} \le M_{\rm He} \le 133 M_{\odot}$);
iii) black hole (BH) collapse after the final explosion 
($9\, M_{\odot} \le M_{\rm He} < 64 M_{\odot}$)
iv) SNII explosion with stellar remnant 
($1.3\, M_{\odot} \le M_{\rm He} < 9 M_{\odot}$)} 
\label{fig_mcore}
\end{figure}
%%%%%%%%%%%%%%%%%%% FIGURE %%%%%%%%%%%%%%%%%%%%%%%%%%%%%%%%%%%%%%%%%%%

\subsection{Chemical enrichment and primordial $\Delta Y / \Delta Z$}

The extent of helium enrichment, $\Delta Y$,  
supplied by Pop-III stars 
in the very early stages of cosmic chemical evolution, 
is a potentially important issue for its possible cosmologic implications.
For instance, the age of Globular Clusters (GC), 
as inferred from observations,
depends sensitively on the assumed initial helium abundance $Y_{\rm p}$, 
which is commonly thought to sample the primordial Big Bang 
nucleosynthesis (BBN). Typically, an increment of the fractional mass 
abundance  $\Delta Y_{\rm p} = 0.02$ would produce  
an age decrement of roughly $15 \%$, that is 
of the order of 2 Gyr within the range $10-15$ Gyr (see e.g. Shi 1995).
Moreover, the  relative enrichment of helium with respect to that of metals
-- the so-called $\Delta Y / \Delta Z$ --  
produced by Pop-III stars, could largely differ from that
derived -- both theoretically and observationally -- at non-zero 
metallicities, so that the usually adopted linear extrapolation 
down to $Z=0$  to infer the primordial helium abundance  
may lead to misleading results.

Another cosmological issue is related to the nature itself 
of the dark matter present in galactic haloes and clusters.
The standard hot BBN predicts an upper limit 
to the baryonic density parameter, $\Omega_{\rm BBN} < 0.06 h_{50}^{-2}$
(where $h_{50}$ is the Hubble constant in units of 50 km $s^{-1}$ Mpc$^{-1}$; 
see e.g. Kernan \& Krauss 1994), that is already lower than 
the present density parameter
inferred from dynamical estimates of virialized systems, 
$\Omega_{\rm dyn} \sim 0.1-0.2$ (e.g. Dekel 1994). 
This discrepancy could be solved --  without losing, at the same time, the   
consistency between  BBN predictions and the observed elemental abundances 
(like those of deuterium and lithium) at the lowest metallicities --
by invoking that the bulk of dark matter is in a 
non-baryonic, still unknown, form. This is 
the commonly accepted picture.

An alternative scenario might arise from  
the assumption that, instead,  dark matter was made of baryons, 
e.g. in form of the remnants of pre-galactic stars. 
Were this the case, the immediate dramatic consequence 
would be that we should abandon the standard BBN, and ascribe the 
first synthesis of light elements, like deuterium and helium, as well as 
the photon microwave background to another source.
Bond et al. (1983) 
and  Carr et al. (1984) pointed at Pop-III very massive stars 
as potential  candidates, and investigated  
the extreme possibility that these stars synthesised 
a substantial fraction, or even the totality, of primordial helium.
In that case, a related constraint is that the accompanying metal enrichment
does not exceed the upper limits set e.g. by the metallicities
measured in Population I and II stars.

In order to derive an updated picture 
of the aforementioned aspects on the basis of the present results,   
we have calculated the helium and metal enrichments produced by
a population of (very) massive zero-metallicity stars,  with
$10\ M_{\odot} \le M \le 8 \times 10^4\, M_{\odot}$.
No other sources are considered, 
such as the wind yields from 
low- and intermediate-mass stars  in the Asymptotic Giant Branch (AGB) phase
(with $0.7\, M_{\odot} \la M \la 8 \, M_{\odot}$).
In other words, we assume that the primordial initial mass function 
favoured the formation of massive objects, 
a hypothesis that is supported by current
models of primeval cloud fragmentation (e.g. Bromm et al. 1999, 2002; 
Nakamura \& Umemura 2001; Abel et al. 2000).
The choice of $8 \times 10^4\, M_{\odot}$ as maximum mass value  is
motivated by the indications that above this limit 
stars are not expected to experience static nuclear burnings
because of post-Newtonian instabilities, and the chemical enrichment 
from these super-massive objects should be small 
(Fricke 1973, Ober 1979; see also Ober et al. 1983).

In summary, we account for the chemical contributions
from three groups of stars, namely:
i) massive objects (MOs) with masses of 12-40 $M_{\odot}$, that
experience no steady mass loss and undergo SN II explosions;
and ii) VMOs with initial masses of 120-260 $M_{\odot}$,
that may experience mass loss via stellar winds 
and pair-instability SN explosions; and iii) VMOs with initial masses of 
$260 - 8 \times 10^4 M_{\odot}$ that may suffer mass loss during the 
hydrostatic phases, before collapsing into black holes.
As for the MOs, we adopt the supernova yields at $Z=0$ calculated by 
 Woosley \& Weaver (1995). For the VMOs with initial masses
of 120-1000 $M_{\odot}$, we
combine our estimated wind contributions with the supernova yields
(if non zero) by  Heger \& Woosley (2002).
The matching between the  wind and SN yields 
is performed as a function 
of the helium core mass as tabulated in  Heger \& Woosley (2002).
Finally, for stars more massive than 1000 $M_{\odot}$, we 
adopt simple analytical prescriptions to estimate the helium and metal
yields ejected by steady mass loss over the hydrostatic hydrogen-burning phase.
Following Bond et al. (1984) we assume 
the maximum  possible helium yield, which is expressed 
as a function of the initial electron molecular weight 
(their equation 27). The corresponding metal enrichment is obtained
by multiplying an average CNO abundance 
(newly synthesised during the core hydrogen-burning, 
typically $\langle X({\rm CNO})\rangle \approx 10^{-9}-10^{-8}$, 
see Sect.~\ref{ssec_rad}) by the amount of ejected
mass (derived by equating their equations 24 and 27).
With our choices for the primordial hydrogen and helium abundances
($X_0=0.77$, and $Y_0=0.23$ respectively), we obtain the fractional yields
-- normalised to the initial stellar mass -- for objects with
$M > 1000\, M_{\odot}$:

\begin{tabular}{lcl}
$y(^4{\rm He})/M$ & = & 0.17 \\
$y({\rm CNO})/M$  & = & 0.435 $\times \langle X({\rm CNO})\rangle$ \\
\end{tabular}

\noindent 
In our calculations we assume $\langle X({\rm CNO})\rangle = 10^{-8}$.

In a similar manner as in Ober et al. (1983), 
a very simple chemical model
is assumed: stars formed in a single burst from a primeval cloud, 
that is assimilated to a closed box system, 
chemical enrichment is considered instantaneous with the burst and 
homogenised throughout the gas.

The basic free parameters that define our model 
are:
the fraction of the primordial cloud, with original mass $M_0$, 
globally consumed in the burst of star formation,
$f_{\rm s} = M_{\rm s}/M_0$, and the details of primordial IMF
(in number of stars per unit mass interval):
\begin{equation}
\phi(M) = \mathcal{A}  M^{-x}\,\,\,\,\,\,\,\,\,\,\,\,\,\,\,\,
\,\,\, {\rm for}\,\,\,\,\ M_{\rm low} \le M \le M_{\rm up}\,\,, 
\label{eq_imf}
\end{equation} 
where $x$ is the slope of the assumed  power law
($x=2.35$ describes the classical Salpeter relation); 
$M_{\rm low}$, $ M_{\rm up}$ correspond 
to the minimum and maximum values of the 
initial stellar mass, respectively.

The quantity $\mathcal{A}$ in Eq.~(\ref{eq_imf}) must satisfy the condition
\begin{eqnarray}
M_{\rm s} & = & \int_{M_{\rm low}}^{M_{\rm up}} M \phi(M) d M =  
\mathcal{A} \int_{M_{\rm low}}^{M_{\rm up}} M^{1-x} d M \\
& =  &  \mathcal{A} \times \xi
\nonumber 
\end{eqnarray} 
Denoting with 
$\Delta^{\rm W}(M)$, $\Delta^{\rm SN}(M)$
the masses of the ejecta contributed by a star with initial mass $M$
via stellar winds and supernova explosions, respectively, then
the final mass of the cloud, left in form of gas 
after the burst of star formation, is
\begin{eqnarray}
\nonumber 
M_{\rm f} & = & M_0 - M_{\rm s} + \mathcal{A} \int_{M_{\rm low}}^{M_{\rm up}} 
[\Delta^{\rm W}(M) + \Delta^{\rm SN}(M)] M^{-x} d M \\
\label{eq_r} 
& = & M_0 - M_{\rm s} + \mathcal{A} \times \mathcal{R}
\end{eqnarray} 
Analogously, the final mass of the cloud in form of any given element $j$
can derived from
\begin{eqnarray}
M_{j} & =  & (M_0 - M_{\rm s}) X_{j,0} + 
\mathcal{A} \int_{M_{\rm low}}^{M_{\rm up}}
\{y_j^{\rm W}(M) + y_j^{\rm SN}(M) \\ 
\label{eq_rj}
\nonumber
 & + & [\Delta^{\rm W}(M) + \Delta^{\rm SN}(M)]X_{j,0}\} M^{-x} dM \\
\nonumber
& = &(M_0 - M_{\rm s}) X_{j,0} + \mathcal{A} \times \mathcal{R}_j
\end{eqnarray}  
where $X_{j,0}$ denotes the initial abundance of the element $j$, 
$y_j^{\rm W}(M)$ and $y_j^{\rm SN}(M)$ are the corresponding 
yields (in mass units) due 
to stellar winds and SN explosion, respectively.
In Eqs.~(\ref{eq_r}) and (\ref{eq_rj}) the quantities $R$ and $R_{j}$
are a measure of the integrated efficiency of the 
primordial stellar population in terms of restitution to the ISM of gas, 
and elemental abundances, respectively.

\begin{figure}
\resizebox{\hsize}{!}{\includegraphics{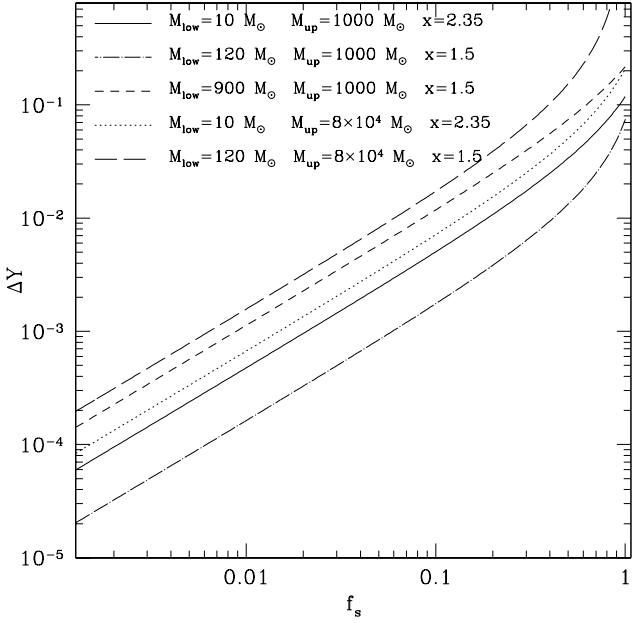}}
\caption{Helium enrichment produced by a primordial simple stellar
population as a function of the star formation efficiency, 
according to different prescriptions, as indicated}
\label{fig_he}
\end{figure}

\begin{figure}
\resizebox{\hsize}{!}{\includegraphics{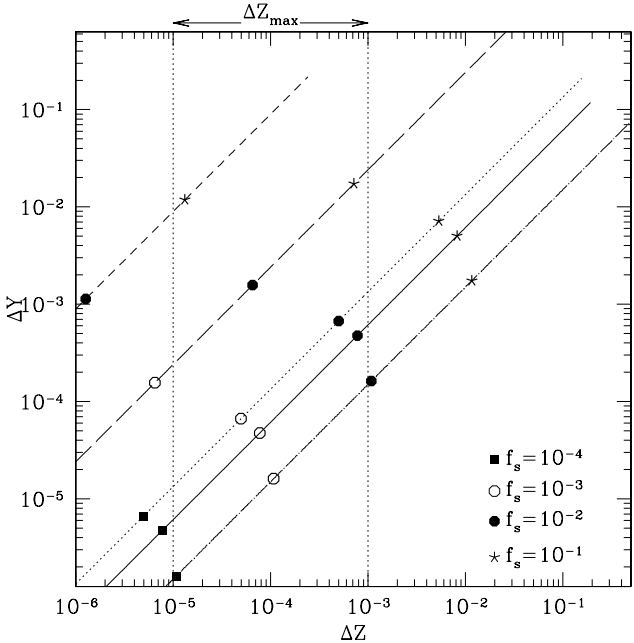}}
\caption{Helium enrichment as a function of  metal production.
The illustrated cases refer to the same prescriptions 
(and corresponding line styles) as described in Fig.~\protect{\ref{fig_he}}.
Selected values of the  star formation efficiency parameter, $f_{\rm s}$, are
marked along the curves}
\label{fig_dydz}
\end{figure}

Finally, after simple algebraic passages, we estimate 
the fractional abundance (by mass) of the element $j$
in the gas resulting from  the chemical enrichment by Pop-III stars:
\begin{equation}
\Delta X_j =  X_{j}-X_{j,0} = \frac{M_{j}}{M_{\rm f}} -X_{j,0}
=  \frac{f_{\rm s}(\mathcal{R}_j-\mathcal{R} X_{j,0})}
{(1-f_{\rm s}) \xi + f_{\rm s} \mathcal{R}} 
%X_{j} = \Delta X_j - X_{j,0} = \frac{M_{j}}{M_{\rm f}} =  
%\frac{(1-f_{\rm s}) \xi X_{j,0} + f_{\rm s} \mathcal{R}_j}
%{(1-f_{\rm s}) \xi + f_{\rm s} \mathcal{R}}
\end{equation}
We apply the above formula to derive the global enrichment of
helium, $\Delta Y$, and metals $\Delta Z$ produced by a Pop-III 
burst of star formation. We set $Y_0=0.23$ and $Z_0=0$.
The results can be analysed as a function of the efficiency of
star formation (expressed by the parameter $f_{\rm s}$), and different
choices of the IMF.
Figures \ref{fig_he} and \ref{fig_dydz} 
show few  illustrative examples, 
that bracket the widest variability range of the results for 
different combinations of the model parameters.
In practice we consider five cases, namely:
i)--ii) classical Salpeter IMF (with $x=2.35$) 
over the mass intervals $10-1000\, M_{\odot}$ and $10-8 \times 10^4\,
 M_{\odot}$; iii)--iv) Salpeter-like but flatter IMF (with $x=1.5$) 
with the inclusion
of VMOs only, over the mass intervals $120-1000\, M_{\odot}$
and $120-8 \times 10^4\, M_{\odot}$; 
and v) the same as the previous cases but for a very narrow mass range
at about $1000\, M_{\odot}$.

%%%%%%%%%%%%%%%%%%% FIGURE %%%%%%%%%%%%%%%%%%%%%%%%%%%%%%%%%%%%%%%%%%%
%%%%%%%%%%%%%%%%%%% FIGURE %%%%%%%%%%%%%%%%%%%%%%%%%%%%%%%%%%%%%%%%%%%

%%%%%%%%%%%%%%%%%%% FIGURE %%%%%%%%%%%%%%%%%%%%%%%%%%%%%%%%%%%%%%%%%%%
%%%%%%%%%%%%%%%%%%% FIGURE %%%%%%%%%%%%%%%%%%%%%%%%%%%%%%%%%%%%%%%%%%%

As for the helium enrichment, illustrated in Fig.~\ref{fig_he},
 we can make the following remarks.
In all cases, $\Delta Y$ increases with the efficiency
of star formation such that an amount of helium, comparable with the 
primordial value predicted by standard BBN, could be produced by Pop-III 
massive stars if $f_{\rm s} \sim 1$. This finding is in
agreement with the results by Bond et al. (1983).
However, leaving aside this extreme and unlikely condition, we
notice in Fig.~\ref{fig_he} that, at any $f_{\rm s}$, 
the largest contributions  come under the assumption of an extremely 
top-heavy IMF either narrowly peaked at $\sim 1000\, M_{\odot}$ 
(short-dashed line), or  
including the contribution of super-massive objects up to 
$8 \times 10^4\, M_{\odot}$ (long-dashed line).

In these particular cases a helium enrichment already of the order of
$\Delta Y \sim 0.01 $, hence able to affect the age determination 
of GCs, may be attained for $f_{\rm s} \approx 0.05 - 0.1$.
Interestingly, these are typical values of the
star formation efficiency 
as indicated by hydrodynamical models of structure and galaxy formation 
(e.g. Abel et al. 1999; Kawata 2001; 
Chiosi \& Carraro 2002), and chemical evolution
of the intra-cluster medium (Moretti et al. 2002).
%%%
Although the efficiency of star formation somewhat depends on the 
assumed IMF regulating the energy input from supernova explosions, 
stellar winds, UV flux 
etc, in reality  it is the result of 
a delicate balance among several concurring physical processes such as the 
energy input above, cooling of gas by radiative mechanisms, and 
dynamical conditions establishing when a gas particle of the hydrodynamical 
models is able to form stars, so that the role of the IMF is of minor 
relevance (see Chiosi \& Carraro (2002) for a detailed 
discussion of these topics). An efficiency of star formation
$f_{\rm s} \approx 0.05-0.1$ is a sort of general rule.
%%%%%
For the Salpeter case (solid and dotted lines in Fig.~\ref{fig_he}), 
$\Delta Y \sim 0.01 $ could be reached if assuming a larger star
formation efficiency,  $f_{\rm s} \approx 0.1-0.2$.

However, before drawing any conclusion, it is necessary to perform
another important check, related to the concomitant production 
of heavy metals, $\Delta Z$. In fact, Pop-III metal enrichment 
should be constrained not to exceed observed upper limits, such as  
the lowest metallicities measured in Population-II Halo stars,
or alternatively the metallicities measured in high-redshift,
possibly little evolved, environments like the Lyman-$\alpha$ systems.

From spectroscopic analyses of Pop-II stars 
we can reasonably take $Z \approx 10^{-5}$ as representative upper limit 
(see e.g. Norris et al. 1996). As for the high-redshift information 
the situation is more complex as reviewed by Pettini (1999, 2000).
At redshift $z \sim 3$,  the estimated metallicity goes from 
$\sim 0.3 -0.1 Z_{\odot}$ in  Lyman break galaxies (LBG), 
$\sim 0.1-0.01 Z_{\odot}$ in Damped Lyman-$\alpha$ systems (DLA),  
down to $0.01-0.001 Z_{\odot}$ in Lyman-$\alpha$ forests.
The metallicity values measured in LBGs and DLAs  
are relatively high and comparable 
with the metallicities measured in today's Pop-I and Pop-II stars, 
so that these systems are likely
to have already experienced significant chemical evolution.
More plausibly, low-density Lyman-$\alpha$ forests may 
exhibit, instead, the genuine nucleosynthetic signature of Pop-III stars.     

Putting together all these pieces of information, in the present 
discussion we assume that the maximum allowed 
metal enrichment should be confined within 
$10^{-5} \la \Delta Z_{\rm max} \la 10^{-3}$.

From inspecting Fig.~\ref{fig_dydz}, it turns out that for the Salpeter case 
(solid and dotted lines) the permitted interval of $\Delta Z_{\rm max}$ 
corresponds to a minor helium enrichment 
$10^{-5} \la \Delta Y  \la {\rm few}\, 10^{-3}$, which is consistent with 
a quite low star formation efficiency, 
in the range $10^{-4} \la f_{\rm s} < 10^{-2}$.
This result agrees with previous findings by Ober et al. (1983), 
and Abia et al. (2001).

On the other hand, in the case of the IMF
peaked at $M ~\sim 1000\, M_{\odot}$ (short-dashed line), 
$\Delta Z_{\rm max}$ implies   
a significant helium enrichment, $\Delta Y > 0.01$,  and a much
larger efficiency of star formation, $f_{\rm s} > 0.1$.
An intermediate situation applies to the case of 
the extremely top-heavy IMF extending up to super-massive objects
(long-dashed line).
However, it is worth noticing  that under the hypothesis that 
the first single burst of star formation produced  
just VMOs with $M ~\sim 1000\, M_{\odot}$ or super-massive objects,   
the primordial gas would have been enriched only with CNO elements 
(and possibly some  $\alpha$-elements), without any contribution 
in heavier metals, like the iron-group elements.

Finally, we remark that the helium-to-metal enrichment, 
$\Delta Y/ \Delta Z$, is independent of  the star formation
efficiency, but is affected by the individual stellar yields
and the IMF features.
According to our calculations, the two lower curves of Fig.~\ref{fig_dydz}
correspond to a $\Delta Y/ \Delta Z  \sim 0.15-0.6$, respectively.
These values are quite low 
compared to recent observational determinations based on HII regions
and predictions of chemical evolution models, 
e.g. $\Delta Y/\Delta Z= 1.9 \pm 0.5$ can be considered a representative
value following Peimbert \& Peimbert (2001). 
On the other hand, the highest curve is described by a huge 
helium-to-metal-enrichment ratio, $\Delta Y/\Delta Z \approx 900$, 
because of the exceedingly-large helium stellar yield compared to that
in the form of CNO elements.
 
It is clear that this is an extreme situation and large 
room is allowed for intermediate results in between 
the case of the Salpeter IMF and that of a IMF peaked at very large masses. 
Ours is meant as an explorative test indicating that if the primordial
IMF was quite different from the classical one, the primordial chemical
enrichment may have followed a different path from that one expected 
under standard prescriptions.

\section{Concluding remarks}
\label{sec_concl}

In this work we have discussed the evolutionary properties 
of zero-metallicity very massive objects 
($120\, M_{\odot} \le M \le 1000\, M_{\odot}$), on the basis of
new calculations that extend the work presented in Paper I. 
Stellar isochrones (see Appendix A) for ages from 16~Gyr down to $10^{4}$~yr  
are available at the web-address http://pleiadi.pd astro.it.

In the attempt to estimate     
the possible effects produced by stellar winds from primordial VMOs,
we adopt recent formalisms that describe the role of  
radiation pressure and stellar rotation as mass-loss driving factors. 
The  emerging picture is the following:

\begin{itemize}
\item At extremely low metallicity 
($Z \approx 10^{-6}$), 
the mechanism of line-radiation transfer should be scarcely efficient, except 
for very large masses, say $\ga 750 M_{\odot}$.
We also find that, as long as the mass loss front penetrates into
the chemical profile left by the convective core during the H-burning 
phase, the maximum CNO abundance exposed at the surface does not exceed
$\approx 10^{-9}-10^{-8}$, which are typical values required by the CNO-cycle
operating in stars with original metal-free composition.
Such degree of chemical self-pollution is also too low to trigger 
efficient radiation-driven winds from zero-metallicity massive stars.
Instead, much larger surface enrichment may be attained whenever the nuclear
products of He-burning, like carbon and oxygen -- potentially
able to produce large mass-loss rates -- were  brought up to the 
surface by either the penetrating mass-loss front, or some dredge-up
process. In our computations this occurs only for the $1000\, M_{\odot}$
model.

\item Our calculations also indicate that rotating very massive  stars 
may actually reach  the critical condition defined 
as the $\Omega\Gamma$-limit, which should be 
likely accompanied by large mass-ejection rates (here taken 
as large as $\dot M_{\rm crit} = 10^{-3}$ yr$^{-1}$).
However, the net impact on mass loss 
should be quite limited, given the extremely short time during which 
these critical regimes can be maintained. 
To this respect, we remind these results 
are obtained in the context of a simplified description 
of stellar rotation, and
the reader should consider the remarks expressed in Sect.~\ref{ssec_cautr}.

\item Finally we recall that, 
according to a recent analysis by Baraffe et al. (2001),
another possible mass-loss driving mechanism, 
namely the pulsation instability -- usually 
at work in VMO models with  ``normal'' 
chemical composition -- has been found of modest potential efficiency 
at zero metallicity, at least for masses $M \la 500\, M_{\odot}$.

\end{itemize}

By combining the supernova yields  
available in the literature with our predicted wind contributions, 
we also evaluate, in a simple way, the chemical enrichment produced by a 
primeval  population of (very) massive stars. 
It is found that a significant helium enrichment,
$\Delta Y \sim 0.01$, may be reached by assuming that the primordial 
conditions in the Universe were such that  
only very massive stars could form, with typical masses of 
$\approx 1000\, M_{\odot}$, or extending into the super-massive 
domain (up to $\approx 10^{5} M_{\odot}$).

In the former case the implied efficiency of gas-to-star conversion 
should be of the order of $10 \%$, and the  first metal enrichment 
should be in the form of CNO elements (no iron-group elements).
On the other side, under the assumption of the standard Salpeter IMF,
the resulting $\Delta Y$ is indeed negligible over the 
maximum allowed enrichment  in metals, 
$10^{-5} \la \Delta Z_{\rm max} \la 10^{-3}$. The corresponding star formation
efficiency should be also be quite small, $\approx 0.01-0.1 \%$.
%%%%

In any case, as already pointed out long ago by Bond et al. 
(1983) and Carr (1994), the interesting possibility arises  that the first 
generation of stars could  mask the true  (somewhat  lower) 
primordial abundance of helium as
predicted by the standard Big Bang Nucleosynthesis (e.g. Olive et al. 1997).
The question and its implications have been recently addressed by 
Salvaterra  \& Ferrara (2002) 
on the base of the present results.
%%%%%

\appendix
\section{Isochrones}
	\label{sec_tableisoc}

%%%%%%%%%%%%%%%%%%% FIGURE %%%%%%%%%%%%%%%%%%%%%%%%%%%%%%%%%%%%%%%%%%%
\begin{figure}
\resizebox{\hsize}{!}{\includegraphics{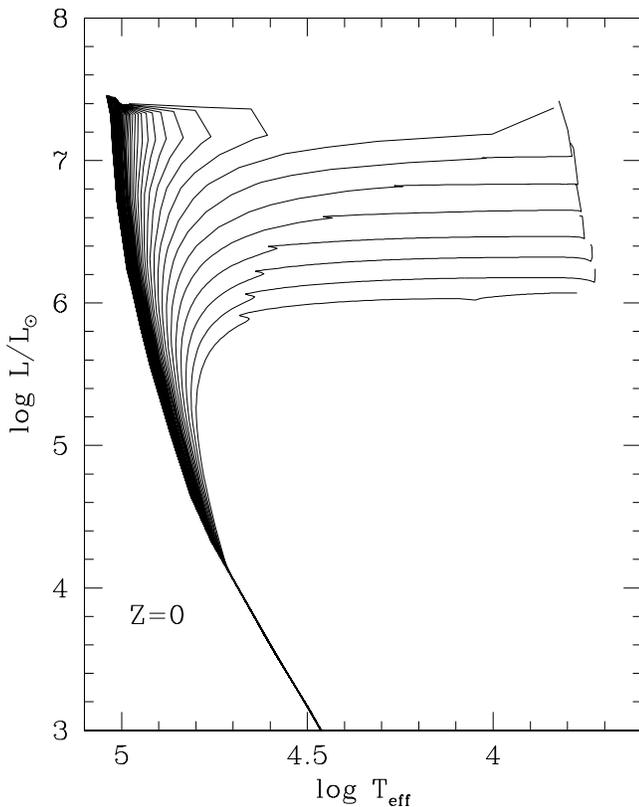}}
\caption{Theoretical isochrones in the HR diagram for the initial 
composition $[Z=0, Y=0.23]$. Age  ranges 
from $\logt=4.0$ to 6.6, at equally-spaced intervals of
$\Delta\log t=0.05$. In all isochrones, the main sequence extends
down to 0.7 $M_{\odot}$ ($\log L/L_{\odot} \sim -0.5$)}
\label{fig_isoc}
\end{figure}
%%%%%%%%%%%%%%%%%%% FIGURE %%%%%%%%%%%%%%%%%%%%%%%%%%%%%%%%%%%%%%%%%%%

Basing on the set of evolutionary tracks for VMOs presented in this work, 
isochrones for very young ages, from $\log t/{\rm yr} =6.6$ down to $4.0$, 
have been constructed.
In this way we complement the 
set of isochrones for zero-metallicity stars 
already presented in Paper I. In fact, 
for ages $\log t/{\rm yr} < 6.6$ the turn-off point corresponds
to stars with $M < 100\, M_{\odot}$, which defines the maximum 
initial stellar  mass in the former set of tracks.
 
An example of the extended isochrones in the H-R diagram is illustrated 
in Fig.~\ref{fig_isoc}.
For ages $2 \times 10^{6} \la t/{\rm yr} \la 10^{8}$ 
the termination point marks the stage of central carbon ignition.

Complete tables with the isochrones can be obtained 
upon request to the authors, or through the WWW site 
http://pleiadi.pd.astro.it. 
They cover the complete age range from about $10^{4}$~yr to 16~Gyr
($4.0\le\log(t/{\rm yr})\le10.25$). Isochrones are provided
at $\Delta\log t=0.05$ intervals.
We refer to Paper I (appendix B) and the quoted web-site for all the details
concerning the description of the table format and layout, 
as well as the sources for magnitude and colour transformations.

%%%%%%%%%%%%%%%%%%%%%%%%%%%%%%%%%%%%%%%%%%%%%%%%%%%%%%%%%%%%%%%%%%%%%%%%%%%
\begin{acknowledgements}
We are grateful to  L\'eo Girardi for his help in the construction
of stellar isochrones; N. Langer, J. Puls,  and N.J. Shaviv  
for useful advice in the matter of stellar winds and stellar 
rotation. We also thank the referee, D. Schaerer, for many important 
remarks that much improved the final version of the paper.
P.M. and C.C. acknowledge financial support from the Italian 
Ministry of Education, University and Research (MIUR).
\end{acknowledgements} 

{}
\end{document}